\documentclass[prd,twocolumn,showpacs,preprintnumbers,superscriptaddress,floatfix,nofootinbib]{revtex4-1}
\usepackage{graphicx,,booktabs}
\usepackage{epstopdf}
\graphicspath{{figures/}}
\usepackage{amssymb}
\usepackage{amsmath,txfonts}
\usepackage{graphicx}
\usepackage{dcolumn}
\usepackage{color}
\usepackage{bm}
\usepackage[colorlinks,
citecolor=blue,
anchorcolor=green,
menucolor=orange,
linkcolor=red,
filecolor=red,
runcolor=pink,
urlcolor=blue,
frenchlinks=red]{hyperref}
\usepackage{multirow}
\usepackage[capitalize]{cleveref}
\usepackage{slashed}
\usepackage[normalem]{ulem}
\makeatletter

\newcommand{\Rmnum}[1]{\expandafter\@slowromancap\romannumeral #1@}

\makeatother

\begin{document}
	
	\title{Understanding of the BESIII measurement of (anti)hyperon-nucleon scattering}
	
	\author{Xiao-Yun Wang}
	\email{xywang@lut.edu.cn}
	\affiliation{Department of physics, Lanzhou University of Technology,
		Lanzhou 730050, China}
	\affiliation{Lanzhou Center for Theoretical Physics, Key Laboratory of Theoretical Physics of Gansu Province, Lanzhou University, Lanzhou, Gansu 730000, China}
	
	\author{Yuan Gao}
	\affiliation{Department of physics, Lanzhou University of Technology,
		Lanzhou 730050, China}
	
	\author{Xiang Liu}
	\email{xiangliu@lzu.edu.cn}
	\affiliation{Lanzhou Center for Theoretical Physics, Key Laboratory of Theoretical Physics of Gansu Province, Lanzhou University, Lanzhou, Gansu 730000, China}
	\affiliation{School of Physical Science and Technology, Lanzhou University, Lanzhou 730000, China}
	\affiliation{Key Laboratory of Quantum Theory and Applications of MoE, Lanzhou University, Lanzhou 730000, China}
	\affiliation{MoE Frontiers Science Center for Rare Isotopes, Lanzhou University, Lanzhou 730000, China}
	\affiliation{Research Center for Hadron and CSR Physics, Lanzhou University and Institute of Modern Physics of CAS, Lanzhou 730000, China}

\begin{abstract}
Focusing on the recent measurements of the scattering processes $\bar{\Lambda} p \rightarrow \bar{\Lambda} p$ and $\Lambda p \rightarrow \Lambda p$ reported by BESIII, we offer a dynamical interpretation of the differences in the measured total and differential cross sections for both processes. This discrepancy arises because the $u$-channel contribution is forbidden for $\bar{\Lambda} p \rightarrow \bar{\Lambda} p$ but allowed for $\Lambda p \rightarrow \Lambda p$. Additionally, we examine the enhancement effect in the forward angle region of the differential cross section for $\bar{\Lambda} p$ elastic scattering, attributed to the $t$-channel contribution. This work serves as a test to the scattering mechanism involved in these reactions.
	\end{abstract}

	\maketitle

Scattering experiments play a crucial role in our understanding of the matter world, revealing the inner structure of atoms and nuclei and facilitating the abundant observation of light flavor hadrons \cite{ParticleDataGroup:2024cfk,wl:2024ext}. With advancements in experimental technology, collider experiments have become the primary means of uncovering the underlying mechanisms of matter. However, it is undeniable that scattering experiments still warrant greater attention.

In 2021, a proposal \cite{Yuan:2021yks} for a scattering experiment was introduced based on the BESIII experiment, which has currently collected 10 billion $J/\psi$ events. The $J/\psi$ meson can decay into antineutrons, hyperons ($\Lambda$, $\Sigma$, and $\Xi$), and their antiparticles, which can interact with protons in the beam pipe material of the BESIII detector, making these typical scattering processes. Generally, obtaining beams of antineutrons, hyperons, and their antiparticles is challenging, and this type of scattering experiment has been rarely explored in the past. The proposal in Ref. \cite{Yuan:2021yks} demonstrates the feasibility of accessing these processes with BESIII. Subsequently, the BESIII Collaboration conducted the first study of the reaction $\Xi^0 n \to \Xi^- p$ \cite{BESIII:2023clq} and measured $\Lambda N$ inelastic scattering \cite{BESIII:2023trh}, providing a preliminary test of this proposal.

Very recently, BESIII reported the first study of antihyperon-nucleon scattering, specifically the processes $\bar{\Lambda} p \to \bar{\Lambda} p$ and the measurement of the $\Lambda p \to \Lambda p$ cross section \cite{BESIII:2024geh}. At first glance, one might expect these two processes to yield similar results, as they differ only by the substitution of $\Lambda$ with $\bar{\Lambda}$. However, the BESIII results reveal a significant discrepancy in the differential cross sections for $\bar{\Lambda} p \to \bar{\Lambda} p$ and $\Lambda p \to \Lambda p$. This situation warrants further investigation through a comprehensive study.

\begin{figure}[htbp]
	\centering
	\includegraphics[width=0.9\linewidth]{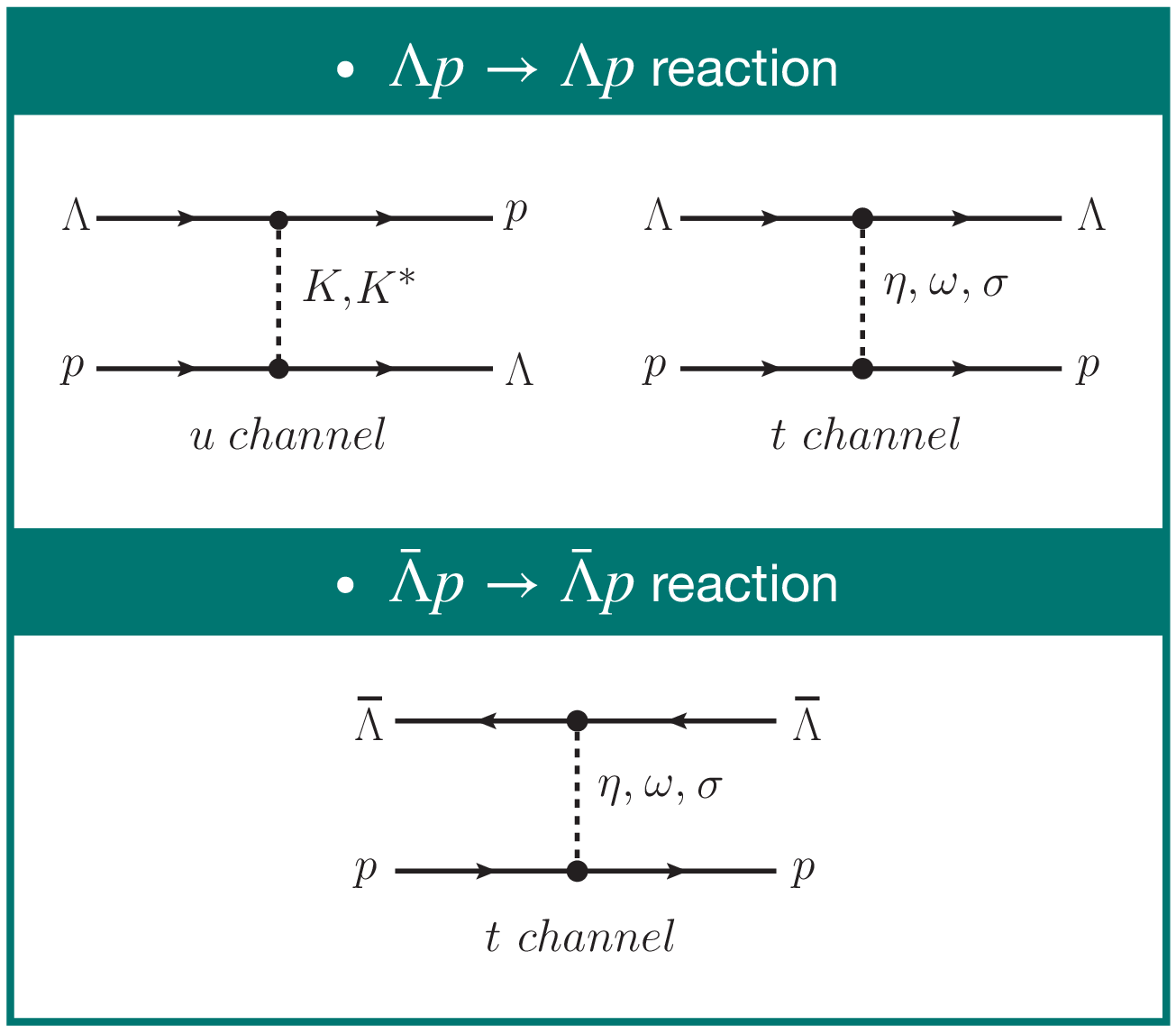}
	\caption{Feynman diagrams for the $\Lambda p\rightarrow\Lambda  p$ and $\bar{\Lambda} p\rightarrow\bar{\Lambda}  p$ reactions.}
	\label{fmtA}
\end{figure}

In this work, we investigate the reactions $\Lambda p \rightarrow \Lambda p$ and $\bar{\Lambda} p \rightarrow \bar{\Lambda} p$ using the one-boson exchange model, as a phenomenological approach, to explore the differences in their differential cross sections. Hadron-hadron scattering processes are generally well-described within the framework of the effective field theory approach \cite{ParticleDataGroup:2024cfk,Bedaque:2002mn}. The one-boson exchange model employed here serves as a simplified version of effective field theory, effectively capturing the primary physical mechanisms. This method has demonstrated successful applications in various scattering processes \cite{Wang:2024qnk,Wang:2023lia,Lin:2022eau,Wang:2022sib,Wang:2019uwk}, motivating its adoption in the present study. Nevertheless, further detailed and comprehensive investigations in this area remain an essential direction for future research.
The tree-level Feynman diagrams for the $\Lambda p \rightarrow \Lambda p$ and $\bar{\Lambda} p \rightarrow \bar{\Lambda} p$ reactions are illustrated in \cref{fmtA}. These diagrams include $u$-channel exchanges of $K$ and $K^*$, as well as $t$-channel exchanges of $\eta$, $\omega$, and $\sigma$. Due to isospin conservation, $\bar{\Lambda}\bar{\Lambda}$ or $\Lambda\Lambda$ cannot couple with isospin-1 mesons like $\pi$ or $\rho$, thus ruling out exchanges of $\pi^{0}$ and $\rho^{0}$ in the $t$-channel. For the $\bar{\Lambda} p \rightarrow \bar{\Lambda} p$ process, the conservation of baryon number at the interaction vertex render the $u$-channel exchange forbidden, allowing only $t$-channel exchanges. Additionally, the $t$-channel contribution may enhance the cross section at forward angles, contributing to the observed differences.
	
	To compute the contributions of the terms shown in \cref{fmtA}, we utilize the interaction Lagrangian densities as outlined in \cite{Wang:2015zcp,Wang:2024qnk,Zhao:2013ffn}. The Lagrangians for the $u$-channel processes are given by:
	\begin{eqnarray}
		\mathcal{L}_{KN\Lambda}&=&-ig_{KN\Lambda}\bar{N}\gamma_5\Lambda K+ \mathrm{H.c.},\\
		\mathcal{L}_{K^*N\Lambda}
		&=&-g_{K^*N\Lambda}\bar{N}\Lambda(K\mkern-11mu/^*-\frac{\kappa_{K^*N{\Lambda} }}{2m_N}\sigma_{\mu\nu}\partial^{\nu} K^{*\mu}) + \mathrm{H.c.}.
	\end{eqnarray}%
	The Lagrangian $\mathcal{L}_{t}$ corresponding to the $t$-channel is expressed as
	\begin{eqnarray}
		\mathcal{L}_{t}
		&=&-ig_{\eta \Lambda\Lambda}\bar{\Lambda}\gamma_{5}\Lambda\eta
		-ig_{\eta NN}\bar{N}\gamma_{5}N\eta\notag\\
		&&-g_{\omega \Lambda\Lambda}\bar{\Lambda}\gamma_{\mu}\omega^\mu \Lambda-g_{\omega NN}\bar{N}\gamma_{\mu}\omega^\mu N\notag\\
		&&+g_{\sigma \Lambda\Lambda}\bar{\Lambda}\sigma \Lambda+g_{\sigma NN}\bar{N}\sigma N,
	\end{eqnarray}%
	where $\eta$, $\omega$, $\sigma$, $K$, and $K^*$ represent the meson fields, while $\Lambda$ and $N$ denote the hyperon and nucleon fields, respectively.
	
	We use the following coupling constants:
 $g_{\eta NN} = 6.34$ \cite{Wu:2023ywu}, $g_{\eta \Lambda\Lambda} = 7.50$, $g_{\omega NN}=15.8$, $g_{\omega \Lambda\Lambda}=10.57$, $g_{\sigma \Lambda\Lambda}= 6.59$ and $g_{\sigma NN}= 9.89$ \cite{Zhao:2013ffn}. The coupling constants $g_{K^*N\Lambda}=-4.26$ and $\kappa_{K^*N\Lambda}=2.66$ are determined using the Nijmegen potential \cite{Stoks:1999bz}. Additionally, the coupling constant $g_{KN\Lambda}$
  is derived from the SU(3) flavor symmetry relation \cite{Oh:2006hm,Oh:2006in,Ozaki:2009wp,Kim:2011rm,Kim:2014hha}
	\begin{eqnarray}
		g_{KN\Lambda} =-\frac{1}{\sqrt{3}}(1+2\beta)g_{\pi NN}=-13.24,
	\end{eqnarray}%
	where $\beta=0.365$ and $g_{\pi NN}^2/4\pi=14.0$. For the vertex involving the anti-hyperon $\bar{\Lambda}$, we adopt the same coupling constants as for the hyperon $\Lambda$.
	
	In our calculations for the
$t$-channel and
$u$-channel, we use the following form factors:
\begin{eqnarray}
\label{form factor}
		F_{t/u}(q) =\mathrm{exp}(-(m^2-q^2)/\Lambda_{t/u}^2),
	\end{eqnarray}%

    The expression for $\Lambda_{t/u}$ is given by $\Lambda_{t/u} = m + \alpha \Lambda_{\text{QCD}}$, where $q$ represents the four-momentum, and $m$ denotes the mass of the exchanged particles. Here, $\Lambda_{\text{QCD}} = 220 , \text{MeV}$. The model parameter $\alpha$ is expected to be of the order of unity \cite{Tornqvist:1993vu,Tornqvist:1993ng,Locher:1993cc,Li:1996yn}, although its precise value cannot be derived from first principles. In this work, the value of $\alpha$ is determined by fitting theoretical predictions to experimental data. For consistency, the parameter $\alpha_t$ is used for $\Lambda_t$ in the $t$-channel, while $\alpha_u$ is used for $\Lambda_u$ in the $u$-channel.
    	
	Based on the Lagrangians described above, the scattering amplitude for the reaction $\Lambda p \rightarrow \Lambda p$ is given by
	\begin{eqnarray}	\mathcal{M}_\Lambda&=&\mathcal{A}_{u}+\mathcal{A}_{t},
	\end{eqnarray}
	while the scattering amplitude for $\bar{\Lambda} p \rightarrow \bar{\Lambda} p$ is
	\begin{eqnarray}	\mathcal{M}_{\bar{\Lambda}}&=&\mathcal{A}_{t},
	\end{eqnarray}
	The reduced amplitudes $\mathcal{A}_{u}$ and $\mathcal{A}_{t}$, corresponding to the $u$- and $t$-channel contributions, are expressed as follows
	\begin{eqnarray}
		\label{Amp1}
		\mathcal{A}^{K}_{u} &=&g_{KN\Lambda}^{2}F_u(q)u(p_2)\gamma_5\bar{u}(p_3)\frac{i}{u-m^2_K}\notag\\
		&&\times u(p_1)\gamma_5\bar u(p_4),\\
		\label{Amp2}
		\mathcal{A}^{K^*}_{u} &=&g_{K^*N\Lambda}^{2}F_u(q)u(p_2)(\gamma_\mu+\frac{\kappa_{K^*N\Lambda}}{4m_N}(\gamma_\mu q\mkern-8mu/_{K^*}-q\mkern-8mu/_{K^*}\gamma_\mu))\notag \\
		&&\times\bar{u}(p_3)\frac{\mathcal{P}^{\mu\nu}}{{u-m_{K^*}^2}}\bar{u}(p_4)
		\notag\\
		&&\times (\gamma_\nu+\frac{\kappa_{K^*N\Lambda}}{4m_N}(\gamma_\nu q\mkern-8mu/_{K^*}-q\mkern-8mu/_{K^*}\gamma_\nu))u(p_1),\\
		\label{Amp3}
		\mathcal{A}^{\eta}_{t} &=&g_{\eta \Lambda\Lambda}g_{\eta NN}F_t(q)u(p_2)\gamma_{5}\bar u(p_4)\frac{i}{{t-m_{\eta}^2}}\notag \\
		&&\times u(p_1)\gamma_{5}\bar{u}(p_3), \\
		\label{Amp4}
		\mathcal{A}^{\omega}_{t} &=&-g_{\omega \Lambda\Lambda}g_{\omega NN}F_t(q)u(p_2)\gamma_{\mu}\bar{u}(p_4)
		\frac{\mathcal{P}^{\mu\nu}}{{u-m_{\omega}^2}}  \notag \\
		&&\times u(p_1)\gamma_{\nu}\bar{u}(p_3), \\
		\mathcal{A}^{\sigma}_{t} &=&-g_{\sigma \Lambda\Lambda}g_{\sigma NN}F_t(q)u(p_2)\bar u(p_4)\frac{i}{{t-m_{\sigma}^2}}u(p_1)\bar{u}(p_3), \notag\\
	\end{eqnarray}%
	where the propagator $\mathcal{P}^{\mu\nu}$ is given by
	\begin{eqnarray}
		\mathcal{P}^{\mu\nu}&=&i(-g^{\mu\nu}+q^\mu q^\nu/{m^2}),
	\end{eqnarray}%
	and the Mandelstam variables are defined as $s=(p_1+p_3)^2$, $u =(p_4- p_1)^2$, and $t =(p_1- p_3)^2$.

Upon completing the necessary preparations, the differential cross section for the reaction $\Lambda(\bar{\Lambda}) p\rightarrow\Lambda(\bar{\Lambda}) p$ can be calculated and used for correlation analysis with experimental data. In the center of mass (c.m.) frame, the differential cross section is expressed as
\begin{eqnarray}
\label{Aamp}
\frac{d\sigma}{dcos\theta} = \frac{1}{32\pi s}\frac{|\vec{p}_3^{c.m.}|}{|\vec{p}_1^{c.m.}|}\left(\frac{1}{4}\sum\limits_\lambda|\mathcal{M}|^2 \right),
\end{eqnarray}%
where $\theta$ denotes the angle between the outgoing $\Lambda$/$\bar{\Lambda}$ baryons and the direction of the incoming $\Lambda$ or $\bar{\Lambda}$ beam
in the center of mass frame. The vectors $\vec{p}_1^{c.m.}$ and $\vec{p}_3^{c.m.}$ represent the three-momenta of the initial $\Lambda$ or $\bar{\Lambda}$ beam and the final $\Lambda$ or $\bar{\Lambda}$ baryons, respectively.

Utilizing the MINUIT code from CERNLIB, we fit the experimental data \cite{BESIII:2024geh} for the reactions $\Lambda p\rightarrow\Lambda p$ and $\bar{\Lambda} p\rightarrow\bar{\Lambda}  p$. Both total and differential cross section data are incorporated into a $\chi^2$ fitting algorithm to determine the values of the free parameters. For the reaction $\Lambda p\rightarrow\Lambda p$, we minimize $\chi^2$ per degree of freedom (d.o.f.) across 10 data points while fitting two parameters: $\alpha_t$ and $\alpha_u$. The fitting parameters, along with a $\chi^2/d.o.f.$ = {\color{red}2.56}, are presented in Table \ref{cutoff}. For the reaction $\bar{\Lambda} p\rightarrow\bar{\Lambda}  p$, we minimize $\chi^2$ per d.o.f. for another set of 10 data points while fitting one parameter:  $\alpha_t$, yielding a $\chi^2/d.o.f.$ =0.90. Table \ref{cutoff} shows that the value of $\alpha_t$ for $\bar{\Lambda} p \rightarrow \bar{\Lambda} p$ is larger than that for $\Lambda p \rightarrow \Lambda p$. This increase in $\alpha_t$ is likely attributed to the higher cross section observed for $\bar{\Lambda} p \rightarrow \bar{\Lambda} p$ at forward angles.


\begin{table}[b]
	\renewcommand\arraystretch{1.2}
	\caption{Fitted values of free parameters based on all experimental data from Ref. \cite{BESIII:2024geh}.}
	\label{cutoff}{\footnotesize \centering
		\setlength{\tabcolsep}{4.5mm}{
			\begin{tabular}{c|c|c|c}
				\hline\hline
				$\Lambda p\rightarrow\Lambda p$& Values&$\bar{\Lambda} p\rightarrow\bar{\Lambda}  p$& Values \\
				\hline
				$\alpha_t$	&-1.42$\pm$0.03 &$\alpha_t$&-1.26$\pm$0.02  \\
				\hline
				$\alpha_u$	&-0.85$\pm$0.15 &&  \\
				\hline
				$\chi^2/d.o.f.$	& 2.56&$\chi^2/d.o.f.$&0.90  \\
				
				\hline\hline
		\end{tabular}}
	}
\end{table}

\begin{figure}[tbp]
	\centering
	\includegraphics[scale=0.43]{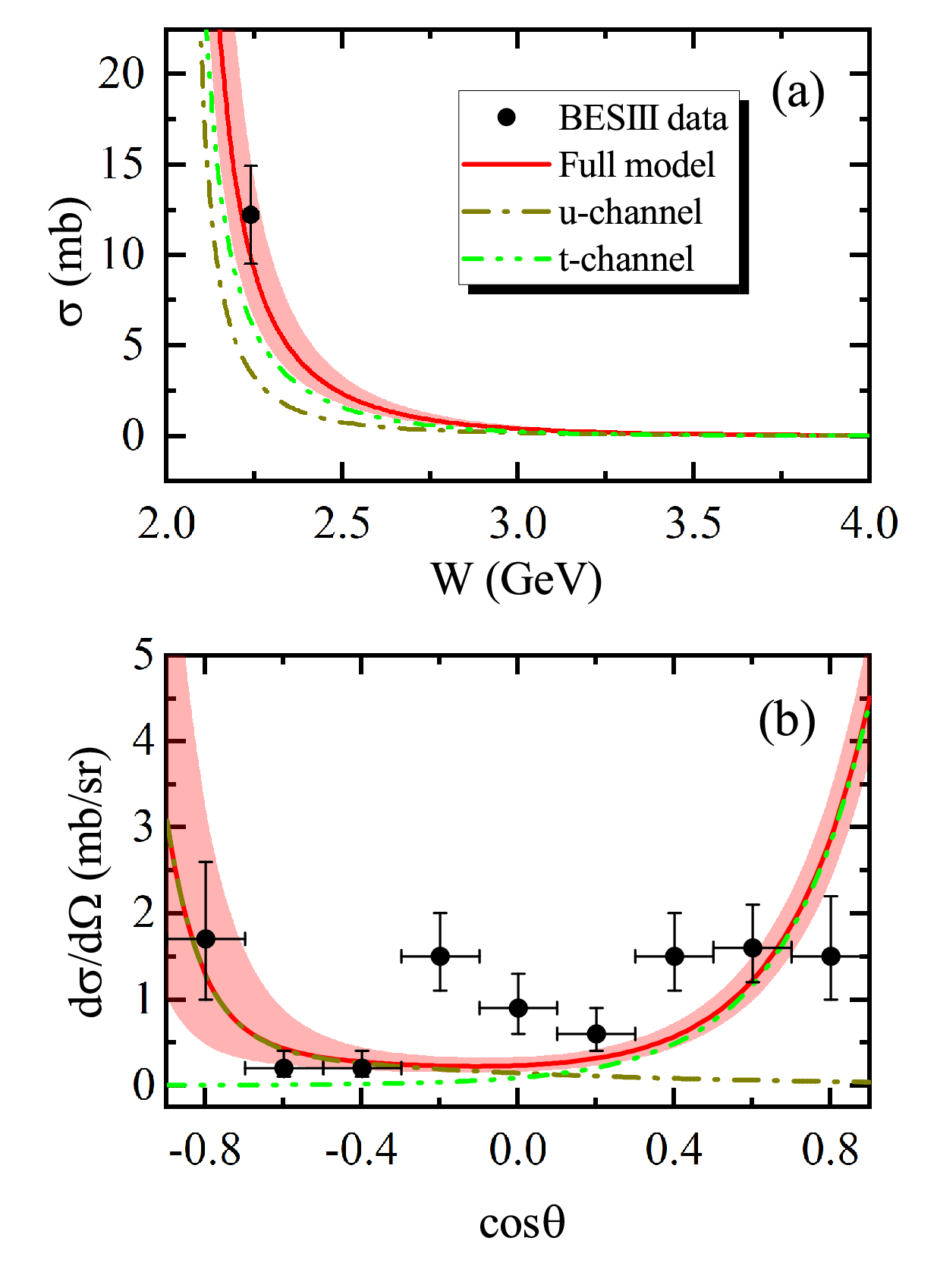}
	\caption{Cross section for the reaction $\Lambda p \rightarrow \Lambda p$. (a) Total cross section; (b) Differential cross section at the center-of-mass (c.m.) energy $W = 2.24$ GeV. Experimental data are from \cite{BESIII:2024geh}. The shaded area represents the error bars for the five fitting parameters in Table \ref{cutoff}. The solid (red), dash-double dotted (green), and dashed-dotted (dark yellow) curves correspond to the full model, $t$-channel, and $u$-channel contributions, respectively.}
	\label{tcsA}
\end{figure}

The total cross section for the reaction $\Lambda p \rightarrow \Lambda p$ is shown in \cref{tcsA} (a), while the differential cross section as a function of $\cos\theta$ at c.m. energy $W=2.24$ GeV
is presented in \cref{tcsA} (b). The results, which incorporate contributions from both the $t$-channel and $u$-channel, demonstrate good agreement with experimental data. The analysis indicates that the $t$-channel contribution dominates at forward angles, whereas the $u$-channel contribution prevails at backward angles.

\begin{figure}[htbp]
	\centering
	\includegraphics[scale=0.43]{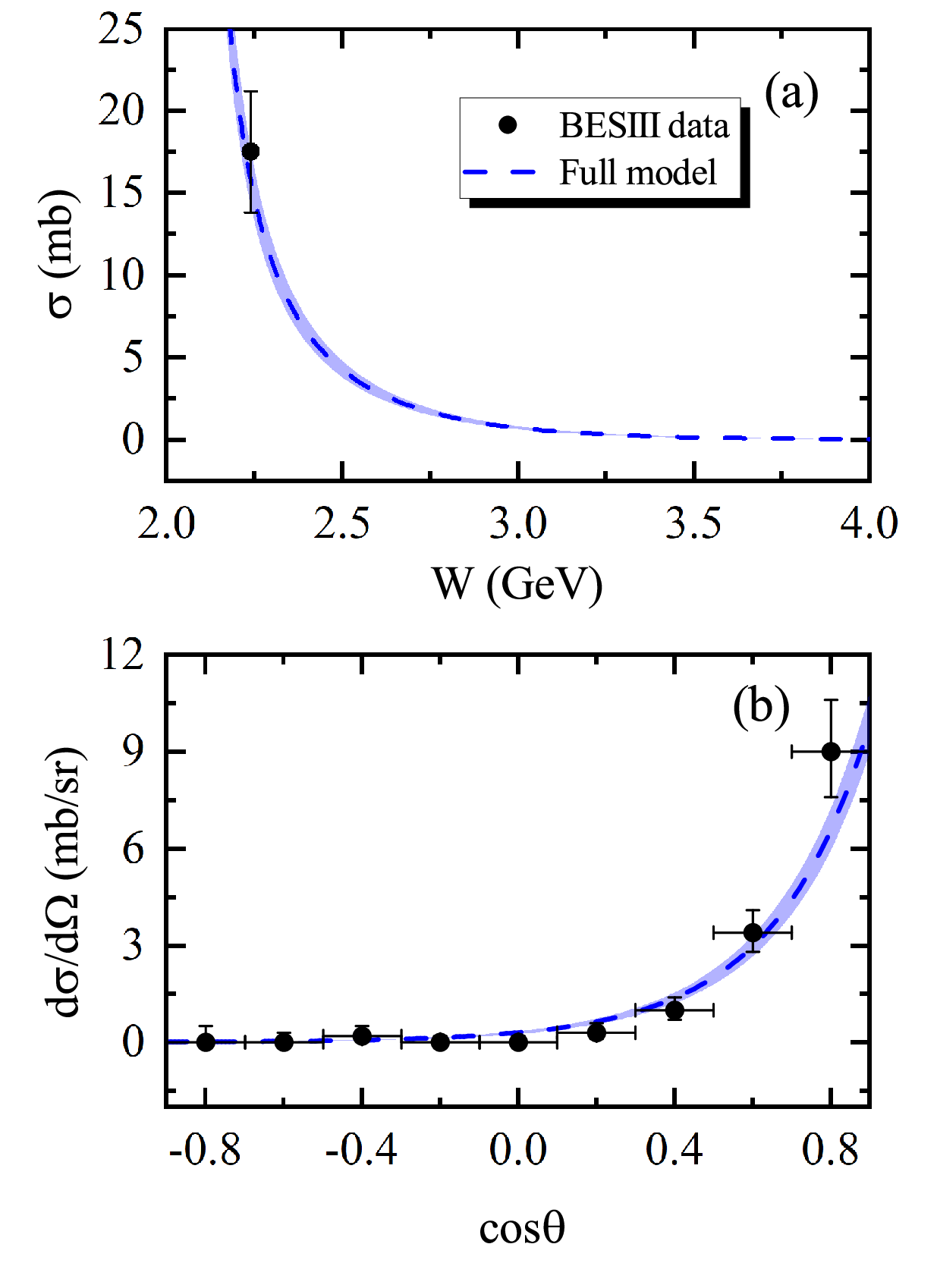}
	\caption{Cross section for the reaction $\bar{\Lambda} p \rightarrow \bar{\Lambda} p$. (a) Total cross section; (b) Differential cross section at c.m. energy $W = 2.24$ GeV. Experimental data are from \cite{BESIII:2024geh}. The shaded area indicates the error bars for the three fitting parameters in Table \ref{cutoff}. The dashed (blue) curve represents the $t$-channel contribution.}
	\label{tcsB}
\end{figure}

\begin{figure}[htbp]
	\centering
	\includegraphics[scale=0.43]{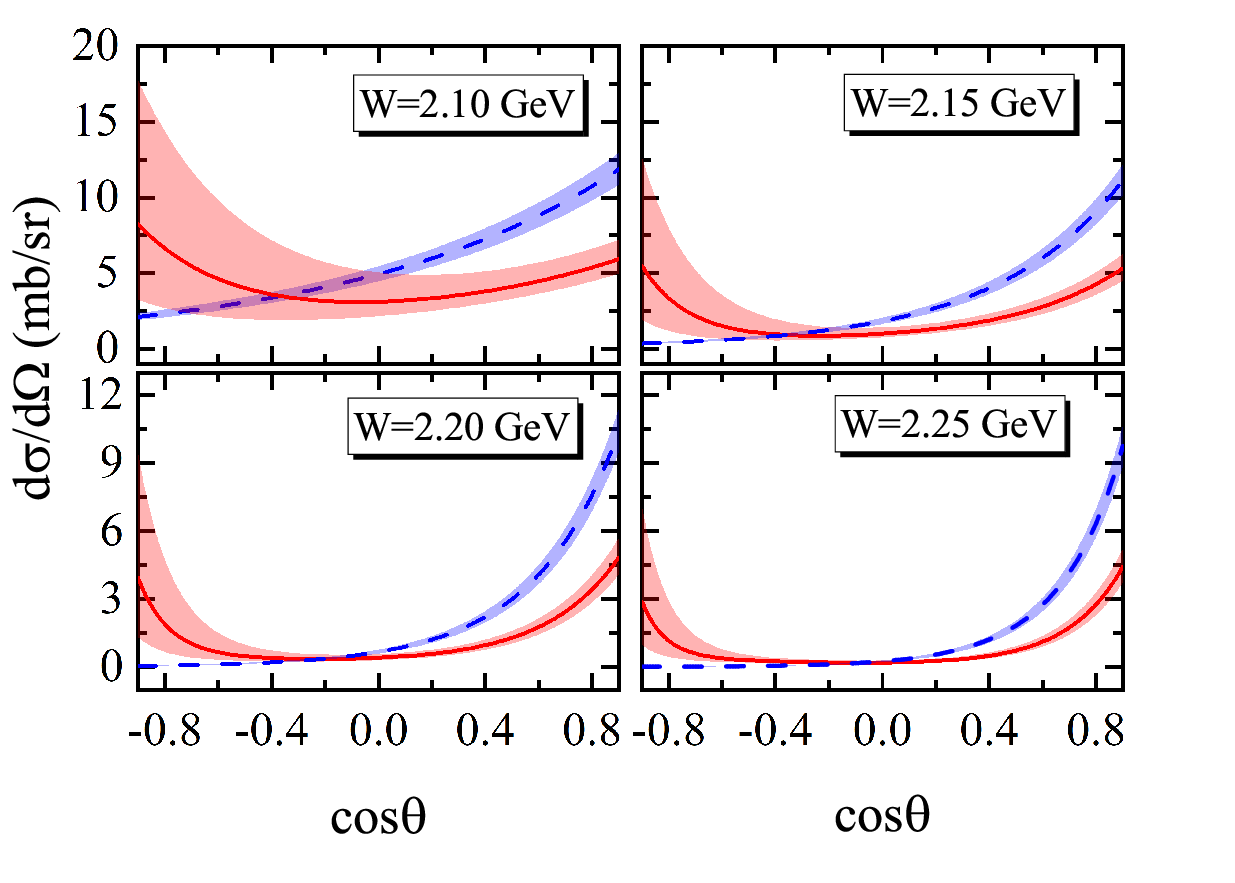}
	\caption{Predicted differential cross section $d\sigma/d\Omega$ for the process $\Lambda(\bar{\Lambda}) p \rightarrow \Lambda(\bar{\Lambda}) p$ as a function of $\cos \theta$ at various c.m. energies. The notation is consistent with that used in \cref{tcsA} and \cref{tcsB}.}
	\label{cosA}
\end{figure}

Following the analysis of the $\Lambda p \rightarrow \Lambda p$ reaction, we also calculate the cross section for $\bar{\Lambda} p \rightarrow \bar{\Lambda} p$, as detailed in \cref{tcsB}. The total cross section for this reaction is shown in \cref{tcsB} (a), while the differential cross section as a function of $\cos\theta$ at c.m. energy $W = 2.24$ GeV is presented in \cref{tcsB} (b). The results, which consider only the $t$-channel contribution, demonstrate good agreement with the experimental data. In \cref{cosA}, we provide four sets of predicted differential cross sections for both the $\Lambda p \rightarrow \Lambda p$ and $\bar{\Lambda} p \rightarrow \bar{\Lambda} p$ reactions at various center-of-mass energies.


In summary, in this paper, we present a detailed analysis of the scattering processes $\Lambda p \to \Lambda p$ and $\bar{\Lambda} p \to \bar{\Lambda} p$, motivated by the significant discrepancy observed in their cross sections, as recently measured by BESIII \cite{BESIII:2024geh}. We provide a dynamical interpretation of these processes, highlighting that the $\Lambda p \to \Lambda p$ reaction includes both $t$-channel and $u$-channel contributions, whereas the $\bar{\Lambda} p \to \bar{\Lambda} p$ reaction involves only the $t$-channel contribution, with the $u$-channel being forbidden. Utilizing an effective Lagrangian approach, we calculate the total and differential cross sections for both reactions and find that our theoretical predictions align well with the experimental data. Thus, the scattering mechanism introduced in this work provides valuable insights into the dynamics of these interactions and enhances our understanding of such scattering processes. We believe that future studies of hyperon-proton scattering at BESIII will yield significant advancements, as the current results discussed here represent a solid foundation. This work serves as an initial step in exploring this important area.

\section*{Acknowledgments}

We would like to thank Xiong-Fei Wang for bringing this new result to our attention.
This work is supported by the National Natural Science Foundation of China under Grants No. 12065014, No. 12047501 and No. 12247101, the Natural Science Foundation of Gansu province under Grant No. 22JR5RA266, and the West Light Foundation of The Chinese
Academy of Sciences under Grant No. 21JR7RA201. X.L. is also supported by National Natural Science Foundation of China under Grant No. 12335001, National Key Research and Development Program of China under Contract No. 2020YFA0406400, the 111 Project under Grant No. B20063, the Fundamental Research Funds for the Central Universities, and the project for
top-notch innovative talents of Gansu province.

\end{document}